\documentclass[3p,twocolumn]{elsarticle}

\usepackage{lineno,hyperref,graphicx,amsmath,amssymb}
\modulolinenumbers[5]

\journal{Physica D}

\begin{document}

\begin{frontmatter}

\title{Simple nonlinear models suggest variable star universality}

\author[hawaii,wooster]{John F. Lindner\corref{cor1}}
\ead{jlindner@wooster.edu}
\ead[url]{physics.wooster.edu/lindner/}
\cortext[cor1]{Corresponding author}

\author[hawaii]{Vivek Kohar}
\author[hawaii]{Behnam Kia}
\author[germany]{Michael Hippke}
\author[hawaii]{John G. Learned}
\author[hawaii]{William L. Ditto}

\address[hawaii]{Department of Physics and Astronomy, University of Hawai`i at M\=anoa, Honolulu, Hawai`i 96822, USA}
\address[wooster]{Physics Department, The College of Wooster, Wooster, Ohio 44691, USA}
\address[germany]{Institute for Data Analysis, Luiter Stra{\ss}e 21b, 47506 Neukirchen-Vluyn, Germany}

\begin{abstract}
Dramatically improved data from observatories like the CoRoT and Kepler spacecraft have recently facilitated nonlinear time series analysis and phenomenological modeling of variable stars, including the search for strange (aka fractal) or chaotic dynamics. We recently argued [Lindner et al., Phys. Rev. Lett. 114 (2015) 054101] that the Kepler data includes ``golden" stars, whose luminosities vary quasiperiodically with two frequencies nearly in the golden ratio, and whose secondary frequencies exhibit power-law scaling with exponent near $-1.5$, suggesting strange nonchaotic dynamics and singular spectra. Here we use a series of phenomenological models to make plausible the connection between golden stars and fractal spectra. We thereby suggest that at least some features of variable star dynamics reflect universal nonlinear phenomena common to even simple systems. 
\end{abstract}

\begin{keyword}
nonlinear dynamics, variable stars, golden ratio, universality
\end{keyword}

\end{frontmatter}


\section{Introduction}

The quality of the best stellar brightness time series has recently improved to the point where sophisticated nonlinear analysis becomes possible. For example, we recently used Kepler spacecraft data to analyze RR Lyrae variable stars~\cite{Lindner, Hippke} and identify a class of ``golden" stars, whose luminosities vary quasiperiodically with two frequencies nearly in the golden ratio, and whose secondary-frequencies exhibit power-law scaling that suggests strange nonchaotic dynamics~\cite{Grebogi, Bondeson, Romeiras}.

\begin{figure}[hbt] 
	\centerline{\includegraphics[width=0.83\linewidth]{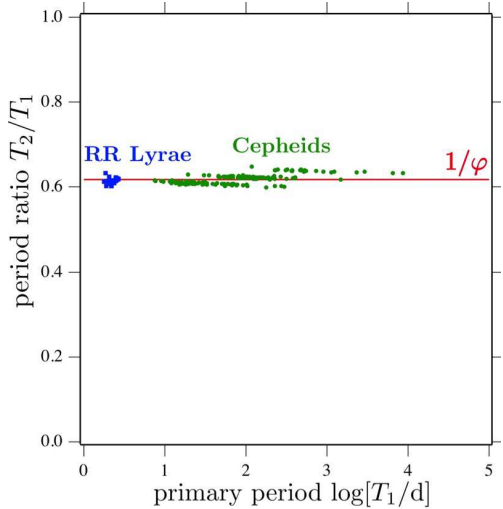}} 
	\caption{Variable star Petersen diagram of period ratio $T_2/T_1$ versus primary period $T_1/\text{d}$ in days has many multifrequency variable stars clustering near the inverse golden ratio $1 / \varphi$. This is a rescaled version of Moskalik's Fig.~2~\cite{Moskalik2013}.	}
	\label{Petersen}
\end{figure}

Many RR Lyrae and Cepheid multifrequency variable stars appear to cluster~\cite{Moskalik2013,Jurcsik,Moskalik2015,Szabo,Netzel} about distinct frequency ratios, including a ratio of approximately 0.62, as in Fig.~\ref{Petersen}. A nonlinear dynamics perspective~\cite{Hilborn} immediately suggests that this is the (inverse) golden ratio $1 / \varphi = 0.618 \ldots$, a frequency ratio that features prominently in quasiperiodic systems and the famous KAM perturbation theorem~\cite{Arnold}. 

Does the multifrequency variable star clustering near the golden ratio result from complicated and particular interactions of higher-order non-radial stellar modes~\cite{Moskalik2015}? Or does it reflect some universal behavior, like the critical exponents that characterize phase transitions, which are common to even simple nonlinear models? Are these stars \textit{accidentally} golden or \textit{fundamentally} golden?

This article presents a series of elementary nonlinear models or analogues of golden stars. These phenomenological models complement ongoing work to develop ab initio hydrodynamic variable star models~\cite{Smolec,Marconi}. Instead of a bottom-up derivation from variable star hydrodynamics to golden-ratio quasiperiodicity and fractal scaling, which may be very difficult, we offer a top-down approach consisting of a sequence of simple models designed to make such a connection credible. The success in nonlinear dynamics of simple models embodying universal features may thereby generalize to variable star astronomy.

We first heuristically motivate the spectral distribution metric used to identify strange nonchaotic dynamics or singular spectra. After summarizing our most recent nonlinear golden star analysis and reviewing some of the relevant and unique  properties of the golden ratio $\varphi$, we describe the models. Simple network models demonstrate how to easily or subtly introduce the golden ratio in stellar caricatures. Potential energy models create stylized versions of golden star spectral distributions or dynamical attractors. An unforced generalized Lorenz flow exhibits golden star scaling. A twist map suggests circumstances in which some dynamical systems can evolve into golden ratio configurations. Finally, we suggest possible new directions for astronomy from nonlinear dynamics.

\section{Spectral Distribution} \label{ScalingSec}

Analysis of the Fourier spectrum of a time series can reveal subtle nonlinear dynamics~\cite{Guzzo}. In particular, strange nonchaotic dynamics \cite{Grebogi, Bondeson, Romeiras} with singular spectra \cite{Pikovsky, Feudel} is dynamics between order and chaos first identified numerically and analytically in quasiperiodically forced systems in the 1980s. The obvious strange nonchaotic signatures of negative maximum Lyapunov exponent and fractal geometry are often difficult to observe, especially in experimental data. However, scale-free power law scaling of a rich frequency spectrum provides a more practical signature, which we have recently discerned in golden stars~\cite{Lindner}. Since this signature is not well known outside the nonlinear dynamics community, we here provide a heuristic derivation of it.

\begin{figure}[htb]
	\centerline{\includegraphics[width=0.75\linewidth]{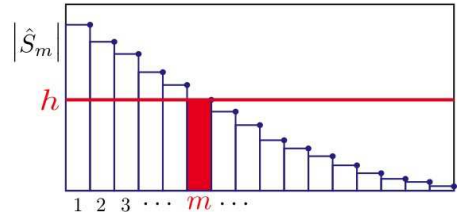} }
	\caption{In a schematic spectrum, $m$ peaks (or bin heights) are higher than the $m$th peak.}
	\label{spectrum}
\end{figure}

Consider a continuous signal $x[t]$ with two prominent incommensurate frequency components $f_1 = 1 / T_1$ and $f_2 = 1 / T_2$. Strobe the signal at the primary period $T_1$ and plot its values versus time modulo the secondary period $T_2$ in the Poincar\'e section
\begin{equation} \label{SectionEq}
	\{t_n, x_n\} =\left\{ n T_1(\bmod\; T_2), x[n T_1] \right\},
\end{equation}
where $n$ are integers. If a function $x_n = S[t_n]$ represents the section, is it smooth? Expand the function as a Fourier series
\begin{equation} \label{FourierEq}
	S[t] = \sum_{m = -\infty}^\infty \hat S_m e^{ i 2 \pi m t / T_2 },
\end{equation}
with derivatives
\begin{equation}
	\partial_t^k S[t] = \sum_{m = -\infty}^\infty \hat S_m (i 2 \pi m / T_2)^k e^{ i 2 \pi m t / T_2 }.
\end{equation}
For smooth sections, expect the Fourier coefficients $\hat S_m$ to decay exponentially, so all the derivatives also decay (as the exponential overwhelms any power). For nonsmooth sections, expect the Fourier coefficients to decay slower, as a power law, so that some derivatives diverge. Specifically, for large Fourier modes $m$, expect
\begin{equation} \label{indexEq}
    	h \equiv \left| \hat S_m\right| \sim 
    	h_0 \left \{ 
        \begin{array}{ll}
             e^{- m / b},  & \text{smooth}, \\
             m^{- 1 / b},   &   \text{nonsmooth},
          \end{array}
    	\right.
\end{equation}
where $b > 0$. Invert to get
\begin{equation} \label{SectralScalingEq}
    	N \equiv m \sim \left \{ 
        \begin{array}{ll}
           	-b \log [ h / h_0 ], & \text{smooth},  \\
             	\left( h / h_0 \right)^{-b},   &   \text{nonsmooth}.
          \end{array}
    	\right.
\end{equation}
Since an averaged spectrum decreases with mode number or frequency, so that $m$ Fourier coefficients are higher than the $m$th coefficient, as in  Fig.~\ref{spectrum}, reinterpret $N \equiv m$ to be the number of spectral peaks higher than the threshold height $h$. Numerical experiments indicate that this relation continues to hold for more general, irregular or sparse spectra. 

The Eq.~(\ref{SectralScalingEq}) power-law spectral distribution $N \sim N_0 h^{-b}$ associated with nonsmooth Poincar\'e sections has been demonstrated in mechanical~\cite{Ditto} and electrical~\cite{Zhou} experiments and appears to characterize golden stars~\cite{Lindner}. The experiments are typically optimized by driving quasiperiodically at two frequencies that are not only incommensurate, but are as incommensurate as possible, so that their ratio is golden (as discussed in Section~\ref{GoldenSection}). The mechanical experiment, for example, obtained good results and power-law scaling by driving near -- but not exactly at -- the golden ratio.

\section{Stellar Analysis} \label{AnalysisSec}

\begin{figure}[htb] 
	\centerline{\includegraphics[width=0.75\linewidth]{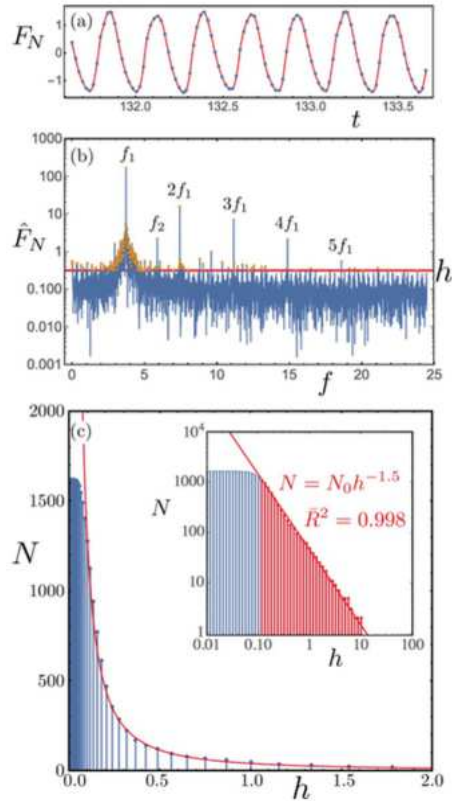} }
	\caption{KIC 5520878 spectral analysis. (a) Excerpt from normalized flux time series $F_N[t]$. (b) Lomb-Scargle periodogram square root $\smash{ \hat{F}_N }$ with primary frequency $f_1$ and secondary frequency $f_2 \approx \varphi f_1$. Gold dots are peaks above example threshold height $h$. (c) Number of superthreshold peaks $N[h]$ exhibits power-law scaling with exponent near $-1.5$ and coefficient of determination $\bar{R}^2 = 0.998$. Inset is log-log plot.}	\label{Analysis}
\end{figure}

As an example of our stellar nonlinear analysis, consider the star of our previous paper, KIC 5520878, whose normalized brightness or flux $F_N$ varies as in Fig.~\ref{Analysis}(a), where Kepler space telescope long cadence data has been conservatively detrended and standardized to zero mean and unit variance~\cite{Lindner}. Although Kepler's time series are of unprecedented quality, they do contain both small and large gaps. Consequently, to estimate the frequency content of the time series, we do not use the Fast Fourier Transform algorithm, which assumes equally spaced points. Rather, we use Least Squares Spectral Analysis~\cite{Press}, effectively fitting sinusoids to the data, by computing the Lomb-Scargle periodogram $\smash{ \hat{F}_N^2 }$ and its square root, as in Fig.~\ref{Analysis}(b).

The primary frequency $f_1$ and its overtones $n f_1$ create the repetitive non-sinusoidal waveform, while the secondary frequency $f_2 \approx \varphi f_1$, a golden ratio higher, modulate the maxima and minima. The spectrum is actually very rich (or very rough); it is not a discrete spectrum of a finite number of peaks, nor is it a continuous spectrum of a smooth curve; rather it suggests a subtle fractal structure, like a singular spectrum. We quantify this by computing the Eq.~(\ref{SectralScalingEq}) spectral distribution of the number $N$ of spectral peaks higher than the threshold $h$, as in Fig.~\ref{Analysis}(c), which results in power-law scaling over two decades in threshold $h$ and three decades in number $N$, with exponent $b \approx 1.5$. Our previous analysis demonstrated similar power-law scaling for the time series sampled (or strobed) at the primary period $T_1 = 1/ f_1$. The current analysis, reported here for the first time, demonstrates similar power-law scaling for the time series itself. Thus, both the original and strobed time series have singular spectra~\cite{Pikovsky}.

Many natural phenomena involve scale-free power-law behavior, at least over a range of variables. For example, the Gutenberg-Richter law implies a number $N \propto A^{-b}$ earthquakes with amplitudes greater than $A$. The exponent $b$ is of order one but is smaller interior to tectonic plates, larger between plates, and even larger for oceanic ridges. Indeed, $b \approx 1.5$ for the volcanic Canary Islands~\cite{Maslov}, so KIC~5520878 stellar flux frequency components scale like Canary Islands earthquake amplitudes. While this quantitative agreement between earthquakes and variable stars is merely an advantageous coincidence, the corresponding intuitions may transfer. Power-law scaling implies a fractional change in one variable is proportional to a fractional change in another. For earthquakes, this might reflect a few larger tremors followed (in time) by many smaller aftershocks. For golden stars, this suggests a few major oscillation modes accompanied (in frequency) by many minor oscillation modes.

\section{Golden Ratio} \label{GoldenSection}

As background for our models, we first describe some of the unique properties of a celebrated mathematical constant, the golden ratio $\varphi$, which is famous in art, physics, and mathematics~\cite{Livio}. For example, the KAM theorem implies that phase space dynamics with two frequencies in the golden ratio maximally resist perturbations~\cite{Arnold,Hilborn}, and nonlinear systems driven by two incommensurate frequencies in an irrational or especially a golden ratio~\cite{Cubero} exhibit distinctive dynamics characterized by strange (aka fractal) \textit{nonchaotic} attractors~\cite{Grebogi, Feudel}. 

If the long and short lengths of a golden rectangle are $\ell$ and $s$, then the golden ratio $\varphi$ of the long to the short is the ratio of the whole to the long, and 
\begin{equation} \label{GoldenEq}
	\varphi = \frac{\ell}{s} = \frac{\ell + s}{\ell} = 1 + \frac{1}{\varphi} = \frac{1 + \sqrt{5}}{2} \approx 1.62,
\end{equation}
which by successive substitutions implies the continued fraction expansion
\begin{equation} \label{GoldenFractionEq}
	\varphi = 1 + \frac{1}{\varphi} = 1 + \cfrac{1}{1+\cfrac{1}{1+\cfrac{1}{{1+\cfrac{1}{1+ \ddots }}}}}.
\end{equation}
All the 1s suggest slow convergence, and indeed $\varphi$ is the irrational number least well approximated by rational numbers. Furthermore, many noble numbers with continued fraction expansions ending in 1s cluster around the golden ratio. By contrast, Liouville's constant
\begin{equation} \label{LiouvilleEq}
	\lambda =\sum_{n=1}^\infty 10^{-n!} = 0.110001000000000000000001\ldots
\end{equation}
is an irrational number very well approximated by rational numbers. This difference in rational approximation or ``nearness" will be important in the models that follow and is why a golden frequency ratio -- and not a Liouville frequency ratio -- is typically used in experiments involving quasiperiodic driving.

\section{Models} \label{ModelsSec}

The following elementary models do not aspire to quantitatively model variable stars. Rather, by exploring the space of possibilities in the simplest possible ways, they complement rather than compete with detailed ab initio hydrodynamic stellar simulations. 

The finite spring network model is an improved version of a model we have previously studied. To the best of our knowledge, the hierarchical spring network model is new. We believe the asymmetric quartic oscillator model is new, but due to its simplicity, it has likely been studied in other contexts. The pressure-countering-gravity model is an elaboration of a model we previously introduced, with the attractor reconstruction being new. Other researchers have studied the autonomous golden flow, but our spectral analysis is new. Twist maps have been studied extensively, including their relation to the KAM theorem, but as far as we know, our numerical experiment summarized by Fig.~\ref{GoldenTwistMap} is new.

\begin{figure}[bt] 
	\centerline{\includegraphics[width=0.5\linewidth]{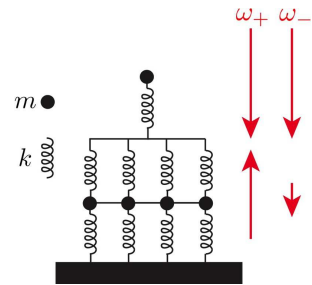} }
	\caption{Finite spring network has two normal modes whose ratio $\omega_{+} / \omega_{-} = \varphi$. Line segments are massless.}
	\label{FiniteSpringNetwork}
\end{figure}

\subsection{Finite Spring Network}

If stars are denser and stiffer in their interiors, then the finite spring network of Fig.~\ref{FiniteSpringNetwork} represents a crude model of a star composed of lumped masses $m$ and ideal springs of stiffness $k$ connected by (horizontal) massless rods. Assuming spring stiffnesses add in parallel and inverse spring stiffnesses add in series, effective spring constants and elementary normal mode analysis implies fast and slow modes with frequencies
\begin{equation}
	\omega_{\pm} = \sqrt{ \left(1 \pm \frac{1}{\sqrt{5}} \right) \frac{k}{m} }
\end{equation}
whose ratio
\begin{equation}
	\frac{ \omega_{+} }{ \omega_{-} } = \frac{1 + \sqrt{5}}{2} = \varphi
\end{equation}
is golden. Because the golden ratio is the root of the Eq.~(\ref{GoldenEq}) quadratic equation, constructing such models is straightforward~\cite{Hippke}.

\begin{figure}[hbt] 
	\centerline{\includegraphics[width=0.7\linewidth]{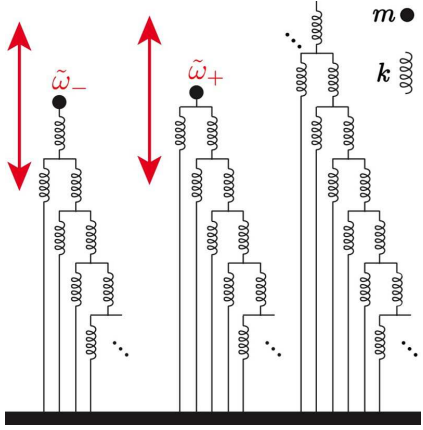} }
	\caption{The two distinct ways (left) to mass terminate the hierarchical spring network (right) oscillate with frequencies whose ratio $\omega_{2} / \omega_{1} = \varphi$. Each node connects to both the bottom layer (like gravity) and the next layer (like pressure).}
	\label{HierarchicalSpringNetwork}
\end{figure}

\subsection{Hierarchical Spring Network}

A more subtle and satisfying model is the hierarchical spring network of Fig.~\ref{HierarchicalSpringNetwork}, where each node connects to both the bottom layer (like gravity) and the next layer (like pressure). Masses terminate the top of the network in two distinct ways, producing effective spring constants
\begin{equation}
	k_{-} = \frac{1}{1/k + 1/k_{-} }
\end{equation}
and
\begin{equation}
	k_{+}  = k + \frac{1 }{1/ k + 1/ k_{+}},
\end{equation}
with oscillation frequencies
\begin{equation}
	\tilde \omega_{\pm} 
	= \sqrt{\frac{k_\pm}{m}}
	= \sqrt{ \left( \frac{\pm1 + \sqrt{5}}{2} \right) \frac{k}{m} },
\end{equation}
whose ratio
\begin{equation} \label{HierarchicalRatoEq}
	\frac{ \tilde\omega_{+} }{ \tilde\omega_{-} } = \frac{1 + \sqrt{5}}{2} = \varphi
\end{equation}
is golden. In practice, the star's center terminates the bottom of the network, truncating the infinite series, so the Eq.~(\ref{HierarchicalRatoEq}) ratio is only approximate, which is conceputally encouraging: ideally the ratio should be exactly $\varphi$ but practically the ratio should be approximately $\varphi$, which is what we observe in golden stars.

\subsection{Asymmetric Quartic Oscillator} \label{AsymmetricSec}

\begin{figure}[bt] 
	\centerline{\includegraphics[width=0.8\linewidth]{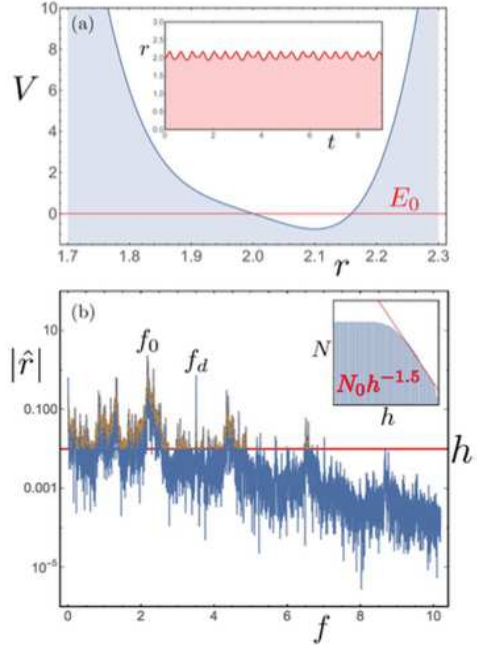} }
	\caption{(a) Asymmetric quartic potential energy with time series as an inset. (b) Spectrum with spectral distribution as an inset.}
	\label{TiltedQuartic}
\end{figure}

A simple continuous model of a golden star is a forced asymmetric quartic oscillator. Combine the Eq.~(\ref{PressureGravityEq}) force law with the potential energy
\begin{equation}
        U = \frac{1}{4} g (r - r_0)^4 - \epsilon (r - r_0),
\end{equation}
as in Fig.~\ref{TiltedQuartic}. Choose the driving frequency to be the golden ratio $\varphi$ times the (numerically computed) natural oscillation frequency $f_0$. The corresponding spectrum is rough and the spectral scaling has a power-law regime with exponent near $-1.5$. In practice, the ``forcing" would be a modulation of one part of the star (say its interior) by another (say its ionized outer layers), as in the ``Eddington valve" or $\kappa$-mechanism~\cite{Eddington} thought to underlie the pulsations of variable stars. Model parameters are $g = 10^4$, $\epsilon = 10$, $r_0 = 2$, and $A  = 5$.

\subsection{Pressure-Countering-Gravity Oscillator} \label{PressureGravitySec}

For a more realistic phenomenological model of a variable star~\cite{Hippke}, recall that a star balances pressure outward versus gravity inward. Assuming spherical symmetry, the radial force on a shell of radius $r$ and mass $m$ surrounding a core of mass $M$ is
\begin{equation} 
	F_r = 4 \pi r^2 P - \frac{G M m}{r^2},
\end{equation}
where the pressure $P/P_0 = \left( V_0/V \right)^\gamma$, and the volume $V/V_0 = \left( r/r_0 \right)^3$. For simplicity, assuming adiabatic compression and expansion and including only translational degrees of freedom, the index $\gamma = 5/3$, and the radial force takes the form
\begin{equation} \label{RadialForceEq}
	F_r =  F_0 \left( \frac{r_e}{r} \right)^2 \left( \frac{r_e}{r} - 1 \right),
\end{equation}
where $r_e$ is the equilibrium radius. The corresponding potential energy
\begin{equation}
	U = U_e \left( \frac{r_e}{r} \right) \left( 2 - \frac{r_e}{r} \right),
\end{equation}
where $U_e = -F_0 r_e /2 < 0$ is the equilibrium potential energy. Oscillate the stellar shell with a frequency a golden ratio times the small oscillation frequency
\begin{equation}
       f_d = \varphi f_0,
\end{equation}
where $2 \pi f_0 = \omega_0 = \sqrt{F_0 / m r_e}$, so that
\begin{equation} \label{PressureGravityEq}
       m  \ddot r = F_r = -\frac{dU}{dr} + A \sin [ 2\pi f_d t ].
\end{equation}

\begin{figure}[bt] 
	\centerline{\includegraphics[width=0.8\linewidth]{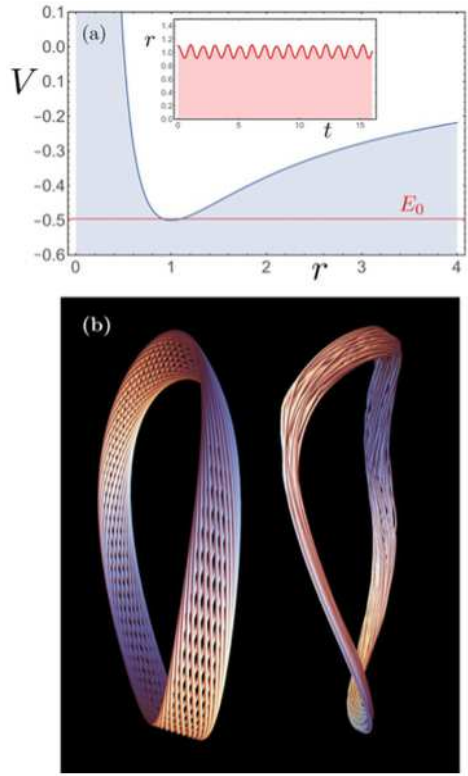} }
	\caption{(a) Pressure-countering-gravity potential energy with time series as an inset. (b) Three-dimensional delay coordinate embedding attractor reconstruction (left) and observed KIC~5520878 attractor reconstruction (right).}
	\label{PressureGravity}
\end{figure}

The resulting stellar flux, which may be proportional to the radius squared or the velocity, is periodic but non-sinusoidal. A typical (numerical) solution is symmetric about the maximum and minimum radii, as in Fig.~\ref{PressureGravity}(a). The curvature at the maxima is (slightly) less than the curvature at the minima due to the difference between the repulsive and attractive contributions to the potential. The three-dimensional delay coordinate embedding attractor reconstruction
\begin{equation}
	 \{r[t], r[t-\tau^\prime], r[t-2\tau^\prime]\}
\end{equation}
on the left of Fig.~\ref{PressureGravity}(b) suggests a stylized version of the observed KIC 5520878 attractor reconstruction
\begin{equation}
	\{F_N[t], F_N[t-\tau], F_N[t-2\tau]\}
\end{equation}
on the right. Model parameters are $m = 1/4\pi^2$, $F_0 = 1$, $r_e = 1$, and $A = 0.015$.

\subsection{Autonomous Golden Flow}

\begin{figure}[htb] 
	\centerline{\includegraphics[width=0.9\linewidth]{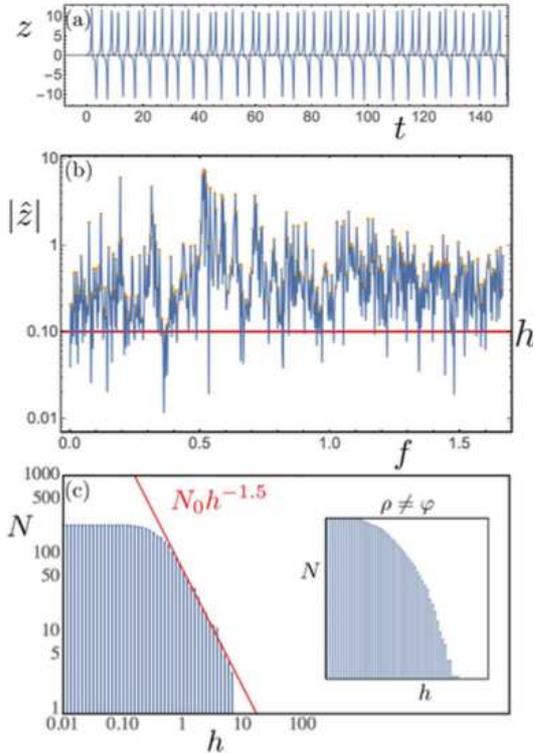} }
	\caption{(a) Autonomous golden flow $z$-component tuned so ratio of maxima to extrema is near a golden rotation number $\rho \approx \varphi$. (b) Corresponding rich  spectrum. (c) Spectral distribution exhibits some power law scaling, which nongolden rotation numbers typically lack, as in the inset for $\rho \neq \varphi$.}
	\label{AutonomousGoldenFlow}
\end{figure}

The asymmetric oscillator produces a rough spectrum with Canary Islands scaling, while the pressure-versus-gravity model generates a plausible time series and attractor, but both are externally forced. An unforced, autonomous model \cite{Lyubimov,Pikovsky} that internally generates the golden ratio and a rough, singular spectrum is a generalized Lorenz convection flow caused by thermal and gravity gradients, as well as vibration, and is described by
\begin{subequations}\label{GoldenFlowEq}
    \begin{align} 
        \dot x &= \sigma y - \sigma D y (z - r), \\
        \dot y &= R x - y - x z,  \\
        \dot z &= x y - b z + a x.
    \end{align}
\end{subequations}
This system produces a butterfly-like attractor similar to but distinct from the canonical Lorenz attractor. The state point alternates between full loops of the butterfly and half loops, as suggested by the $x$-projection of Fig.~\ref{AutonomousGoldenFlow}(a). Indeed the parameters of the model can be adjusted so that the ``rotation number"
\begin{equation}
	\rho= \frac{ \text{half loops} }{\text{full loops}} = \frac{ \text{maxima} }{\text{extrema}} \approx 1.62 \approx \varphi
\end{equation}
is golden. The result is a rough spectrum with a familiar power law scaling, as in Figs.~\ref{AutonomousGoldenFlow}(b) and~\ref{AutonomousGoldenFlow}(c),  at least over a decade in threshold and two decades in number. Model parameters are $R = 14.1487968$, $D = 0.05433476$, $a = -0.56112733$, and the canonical $\sigma = 10$, $b = 8/3$, with sampling time $\delta t = 0.3$.

\subsection{Golden Twist Map}

\begin{figure}[htb] 
	\centerline{\includegraphics[width=1.0\linewidth]{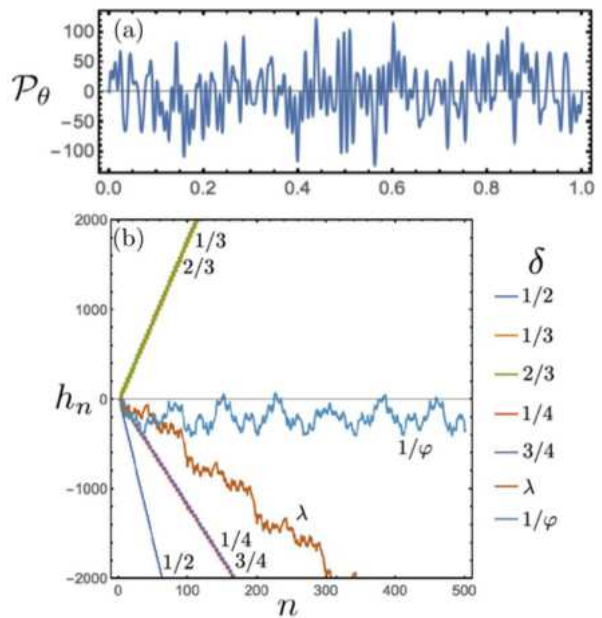} }
	\caption{(a) Twist map ``push" perturbation is very oscillatory. (b) Twist maps diverge up and down cylinder height $h_n$ for rational and near rational shifts $\delta$ leaving the golden shift $\delta = 1 / \varphi$.}
	\label{GoldenTwistMap}
\end{figure}

In nonlinear dynamics, two broad classes of models are flows specified by differential equations and maps specified by difference equations. Our final model is a map that suggests how a golden ratio might survive in a competition among various modes of a variable star. Consider \cite{Ghys} the twist map on a cylinder
\begin{subequations}
    \begin{align} 
    	\theta_n &= \theta_{n-1}+\delta \pmod{1}, \\
	h_n &= h_{n-1}+\mathcal{P}_{n-1},
    \end{align}
\end{subequations}
where $\theta$ is an angle around the cylinder, measured in rotations, $h$ is the height, and $\mathcal{P}$ is a ``push perturbation". If the push $\mathcal{P}$ is identically zero, the twist map becomes a circle map, and all orbits are at the same height, filling the circle densely if the shift $\delta$ is irrational. If the push $\mathcal{P}$ is nonzero on average, the orbits will run away up or down the cylinder, so instead assume a zero average push $\langle \mathcal{P}_\theta \rangle = 0$. In particular, assume the push is the very wiggly sinusoidal sum
\begin{equation}
	\mathcal{P}_\theta = \epsilon \sum^N_{n=1} c_n \sin[n 2 \pi \theta],
\end{equation}
where the coefficients $c_n \in \{ -N, -N+1, \ldots N-1, N \}$ are fixed but randomly distributed, as in Fig.~\ref{GoldenTwistMap}(a).

Iterate the map under very high numerical precision (up to 1000s of digits) for different shifts $\delta$, as in Fig.~\ref{GoldenTwistMap}(b). Simple rational shifts ``resonate" with the push perturbation and run away: one-half escapes rapidly downward, the thirds follow upward, and then the quarters downward. The Eq.~(\ref{LiouvilleEq}) Liouville's constant, which is irrational but easily approximated by rationals, resonates weakly with the push and wanders away more slowly. However, the (inverse) golden shift $\delta = 1/\varphi$, which is the irrational least well approximated by rationals, does not resonate with the push perturbation and stays nearby. For an ensemble of such maps, only the golden shift map endures.

In a continuous system of many interacting modes, like a variable star, if the resonance of a primary mode with a secondary mode leads to large amplitudes and large dissipation, with better resonance corresponding to proportionally larger effects, than the least resonant mode, a golden ratio from the primary mode, might be the last survivor. An engineering example would be circular particle accelerators where resonance between ``betatron" oscillations (\textit{about} the circular orbit) and revolutions (\textit{around} the circular orbit) can eject particles out of the beam and into collisions with the beam walls, forcing operators to routinely avoid simple or ``vulgar" ratios of oscillations to revolutions.

The golden twist map might seem a counter example to the controversial Molchanov hypothesis \cite{Molchanov}, which avers that every oscillatory system evolves to a resonance governed by a family of integers, like the $3:2$ resonance between the orbits of Pluto and Neptune, the $4:2:1$ resonance among Jupiter's giant natural satellites Ganymede and Europa and Io, and the many $1:1$ resonance between tidally-locked moons and their planets. These solar system examples probably formed via non-Hamiltonian viscosity and may be a different class of phenomena. Indeed, overlapping resonances may account for evidence of \textit{chaos} in the solar system~\cite{Lecar}. Furthermore, in the current context of approximately golden stars, it is worth noting that such resonances are not as simple as they seem~\cite{Lainey, Greenberg}; for example, Europa's period is only approximately twice Io's due to the precession of Io's orbit, and it appears to be evolving.

\section{Discussion} \label{DiscussionSec}

Some natural dynamical patterns result from universal features common to even simple models. Other patterns are peculiar to particular physical details. Is the frequency distribution of variable stars universal or particular?

The Feigenbaum constant $\delta \approx 4.67,$ which characterizes the period doubling route to chaos~\cite{Feigenbaum}, has been observed in many diverse experiments. Does the golden ratio $\varphi \approx 1.62$ (or equivalently the inverse golden ratio $1/\varphi \approx 0.62$) play a similar role? Or does the mysterious factor of $0.62$ that characterizes many multifrequency stars result merely from higher-order nonradial stellar oscillation modes?

The simple nonlinear models of this paper suggest the importance of considering simple explanations. The golden ratio itself has unique and remarkable properties; as the irrational number least well approximated by rational numbers, it is the least ``resonant" number. A finite network model of identical springs and masses has two normal modes whose frequency ratio is golden. An infinite network hierarchy can be mass terminated in two ways to naturally generate two modes whose frequency ratio is golden, while a realistic truncation of the model generates a ratio near golden, as observed in the golden stars. A simple asymmetric nonlinear oscillator produces a rich spectrum with a power-law spectral distribution. A more realistic oscillator model of pressure countering gravity exhibits a recognizable but stylized golden star dynamical attractor. An unforced Lorenz-like convection flow also produces a singular spectrum with a power-law spectral distribution, provided its parameters are tuned so that a golden ratio characterizes its orbit. Finally, an ensemble of twist maps naturally evolve to a golden state, because golden shifts are least resonant with any oscillatory perturbation.
 
Helioseismology and asteroseismology have observed many seismic spectral peaks in the sun and other nonvariable stars, which correspond to thousands of normal modes. Yet, despite preliminary analysis, we have not discovered power law scaling in the solar oscillation spectrum. Stochasticity and turbulence dominate the pressure waves in the sun that produce its standing wave normal modes. In contrast, a varying opacity feedback mechanism~\cite{Eddington} inside a variable star creates its regular pulsations. In golden stars, interactions with this pulsating mode may dissipate all other modes except those a golden ratio away.

Simple ratios do describe some astrophysical resonances, but golden ratios may describe others. Gravity is nonlinear, and nonlinearity is a prerequisite for chaos. Future exoplanet spacecraft like TESS and PLATO will greatly refine our knowledge of variable stars and soon test our universality thesis. It would not be surprising to find chaos among the stars.

\section*{Acknowledgments} 

We gratefully acknowledge support from the Office of Naval Research under Grant No. N00014-12-1-0026 and STTR grant No. N00014-14-C-0033. J.F.L. thanks The College of Wooster for making possible his sabbatical at the University of Hawai'i at M\=anoa. We thank J. D. Bjorken for helpful discussions.

\section*{References}

\bibliography{mybibfile}

\end{document}